\documentclass[mathleft]{an}
\usepackage{graphicx}
\usepackage{times}
\begin{document}
%\clubpenalty10000
\sloppy
\def\kcet{$\kappa^1$~Ceti } 
% The following seven commands are intended for editorial usage and should be ignored by
% the author(s).
\Pagespan{000}{000}% Document's page range. 
% If second parameter is left empty, the last page is computed automatically.
\Yearpublication{2010}%
\Yearsubmission{2010}%
\Month{00}%   
\Volume{0000}%  
\Issue{00}% 
% \DOI{This.is/not.aDOI}% 

\title{Automated Asteroseismic Analysis of Solar-type Stars}

\author{C. Karoff\inst{1,2}\fnmsep\thanks{\email{karoff@bison.ph.bham.ac.uk}}, T.~L.~Campante\inst{2,3}, W.~J.~Chaplin\inst{1}}
%\newline}, 
%T.~L.~Campante\inst{2}, P.-O.~Quirion\inst{3} \and W.~J.~Chaplin\inst{1}
%}
\titlerunning{Automated Asteroseismic Analysis of Solar-like Stars}
\authorrunning{C. Karoff}
\institute{
School of Physics and Astronomy, University of Birmingham, 
Edgbaston, Birmingham B15 2TT, UK
\and
Department of Physics and Astronomy, Aarhus University, Ny Munkegade 120, DK-8000 Aarhus C, Denmark
\and
Centro de Astrof\'isica, Faculdade de Ci\^encias, Universidade do Porto, Rua das Estrelas, 4150-762 Porto, Portugal
}

\received{??}
\accepted{??}
\publonline{later}

\keywords{methods: data analysis -- stars: late-type -- stars: oscillations}

\abstract{The rapidly increasing volume of asteroseismic observations on solar-type stars has revealed a need for automated analysis tools. The reason for this is not only that individual analyses of single stars are rather time consuming, but more importantly that these large volumes of observations open the possibility to do population studies on large samples of stars and such population studies demand a consistent analysis. By consistent analysis we understand an analysis that can be performed without the need to make any subjective choices on e.g. mode identification and an analysis where the uncertainties are calculated in a  consistent way.\\\\
Here we present a set of automated asterosesimic analysis tools. The main engine of these set of tools is an algorithm for modelling the autocovariance spectra of the stellar acoustic spectra allowing us to measure not only the frequency of maximum power and the large frequency separation, but also the small frequency separation and potentially the mean rotational rate and the inclination. \\\\
The measured large and small frequency separations and the frequency of maximum power are used as input to an algorithm that estimates fundamental stellar parameters such as mass, radius, luminosity, effective temperature, surface gravity and age based on grid modeling. \\\\
All the tools take into account the window function of the observations which means that they work equally well for space-based photometry observations from e.g. the NASA {\it Kepler} satellite and ground-based velocity observations from e.g. the ESO {\it HARPS} spectrograph.
}
\maketitle
\section{Introduction}
Asteroseismology is proving itself as a powerful tool to increase our understanding of stars. This progress is down due to both ground-based facilities such as {\it Elodie} (Baranne et al. 1996), {\it CORALIE} (Queloz et al. 2000), {\it HARPS} (Mayor et al. 2003), {\it UCLES} (Tinney et al.2001) and {\it UVES} (Dekker et al. 2000) and space-based facilities such as {\it WIRE} (Buzasi et al. 2000), {\it MOST} (Walker et al. 2003), {\it COROT} (Appourchaux et al. 2008) and {\it Kepler} (Chaplin et al. 2010). And we do not expect the fun to stop here. For the future we have high expectations regarding the ground-based telescope network {\it SONG}  (Grundahl et al. 2009) and the ESA satellite {\it PLATO} (Catala 2008).

Solar-type stars are, due to their rich and structural oscillations spectra some of the best targets for asteroseismology. From the characteristic large frequency separation between oscillation modes with different radial order we can get a direct measurement of the (acoustic) radius of the stars (Christensen-Dalsgaard et al. 2007) and by combining this measurement with the characteristic small frequency separation between modes with different angular degree we can get estimates of the mass and the age of the stars (Christensen-Dalsgaard et al. 2007). The acoustic background that we measure in the stars provides us with interesting information about activity and convection (Karoff 2008). For the best observations it is possible to measure the frequencies of the individual oscillation modes which provide us with an inexhaustible source of information on stellar structure and evolution -- including: information on energy generation and transport, rotation and stellar cycles (Karoff et al. 2009).
\section{Methodology}
In this paper we present a set of tools that can be used in an automated fashion to perform an asteroseismic analysis. The tools consist of the following four main parts: 
\begin{enumerate}
\item[I]  Measure the large frequency separation
\item[II] Measure parameters in the acoustic background
\item[III]  Model the autocovariance spectrum ($\Delta \nu, \delta \nu, \nu_{\rm max}$)
\item[IV]  Measure stellar fundamental parameters
\end{enumerate}
The first step performs a search for the characteristic frequency separation and eventually measures the frequency of this large frequency separation ($\Delta \nu$) and the frequency range in which the oscillations modes show a significant characteristic spacing ($\nu_{\rm max}$). The second step uses these first values of $\Delta \nu$ and $\nu_{\rm max}$ to measure the amplitude of the oscillations. This is done using the method described by Kjeldsen et al. (2008). In this step we also obtain measurements of the characteristic time-scale and amplitudes of different components in the acoustic background. The third step uses the newly developed technique for modeling the autocovariance spectrum of the acoustic spectrum by Campante et al. (2010) to obtain more reliable estimates of not only $\Delta \nu$ and $\nu_{\rm max}$, but also the small frequency separation $\delta \nu$ and in the best cases the rotation rate and inclination. All these parameters are not only measured in a consistent manner using the technique developed by Campante et al. (2010) they also come with consistent uncertainty estimates. In the last and fourth step we use $\Delta \nu$, $\nu_{\rm max}$ and $\delta \nu$ plus parameters from the Kepler Input Catalog (Latham et al. 2005) as input to the SEEK routine developed by P.-O.~Quirion in order to obtain a consistent set of stellar parameters for these stars. 
\begin{figure}
\centering
\includegraphics[width=5cm]{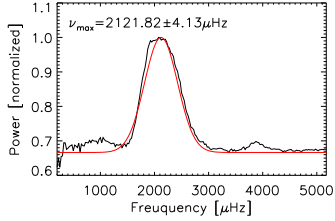}
\caption{Maximum summed power as a function of frequency for the simulation of KIC~10454133. The spectrum shows a clear peak just above 2000 $\mu$Hz reflecting $\nu_{\rm max} = 2121.82 \pm 4.13 \mu$  Hz. The red line shows a fit to the spectrum based on a model that includes an exponential background and a Gaussian peak.} 
\end{figure}

In contrast to most other automated asteroseismic packages (Hekker et al. 2010; Huber et al. 2009; Mathur et al. 2009) the tools presented here all use amplitude spectra calculated using least squares sine wave fitting (Karoff 2008). Generally it only takes of the order of minutes to calculate the spectra used in the analysis presented here.   

The tools presented are automated in the sense that they can run from the beginning to the end without human interaction, but we do demand that all results are verified by a human before a detection is claimed. I.e. if the human eye cannot see any significant signal in any of the plots presented below a detection is not claimed. We have tested this approach on artificial data in the AsteroFLAG framework with very positive results (Chaplin et al. 2008). 

\section{The large frequency separation}
For the Kepler observations we search for a characteristic frequency spacing between 5 and 400 $\mu$Hz in the frequency interval 20 to 7000 $\mu$Hz. This is done by calculating a correlation matrix of summed power spectra:
\begin{equation}
C(\Delta \nu, \nu_{\rm max})=\sum_{\nu=\nu_{\rm max}-2.5\Delta\nu}^{\nu=\nu_{\rm max}+2.5\Delta\nu}{P({\nu})}.
\end{equation}
Here we test 5 radial orders. Of course oscillation modes over more radial orders will be excited in many stars, but in this first step we measure only the characteristic spacing between the 5 most prominent of them. We start by looking for a frequency range that show signs of a characteristic spacing (see Fig.~1). The mean frequency, the width and the uncertainty of this frequency range are all found by fitting a Gaussian to the one-dimensional correlation matrix as a function of frequency, using least squares (see Fig.~1). 

Having a measure of $\nu_{\rm max}$ we then measure the characteristic frequency spacing, i.e. half the large frequency separation ($\Delta \nu$). This is done the same way as for $\nu_{\rm max}$, i.e. by fitting a Gaussian to the one-dimensional correlation matrix as a function of the large separation using least squares (see Fig. ~ 2).
\begin{figure}
\centering
\includegraphics[width=5cm]{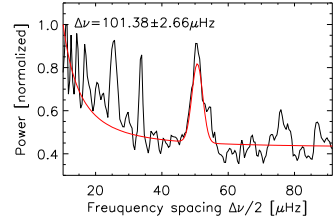}
\caption{Maximum summed power as a function of frequency lag for the simulation of KIC~10454133. The spectrum shows a clear peak just above 50 $\mu$Hz reflecting $\Delta \nu = 101.38 \pm 2.66 \mu$  Hz. The red line shows a fit to the spectrum based on a model that includes an exponential background and a Gaussian peak.} 
\end{figure}
\section{Parameters in the acoustic background}
The charateristic time-scale and amplitude of the different components in the acoustic spectra are measured using a Harvey-like model (Harvey 1985):
\begin{equation}
B(\nu)=\sum_i{\frac{4\sigma^2_i\tau_i}{1+(2\pi\nu\tau_i)^{\alpha}}}+\sum_j{\sigma_{j}e^{\frac{\left(\nu-1/\tau_j^2\right)}{2w_j^2}}}+c,
\end{equation}
where $\sigma$ is the amplitude of the component and $\tau$ is the characteristic time-scale. In intensity we generally see activity, granulation, faculae, p-mode and chromospheric oscillations in the Sun and other solar-type stars (Karoff, 2008). Of course only activity, granulation and faculae will act according to the Harvey-like model in the acoustic spectrum, whereas we assume that the p-mode and chromospheric oscillations show a Gaussian-like signature in the acoustic spectrum (Lefebvre et al. 2008). The spectra are smoothed by a Gaussian running mean with a width equaling two times $\Delta \nu$ before the amplitudes and characteristic time-scales are measured. The reason for this is twofold: firstly it ensures that we can see components such as faculae with small amplitudes compared to the noise and secondly it allows us to assume that the noise in the spectra is normally distributed so we can use least squares to measure the amplitudes and characteristic time-scales.

\section{The autocovariance spectrum of the stellar acoustic spectrum}
\begin{figure}
\centering
\includegraphics[width=5cm]{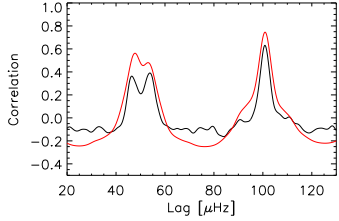}
\caption{Autocovariance spectrum of the acoustic spectrum of simulation of KIC~10454133 overlaid by the model in red. The spectrum shows prominent features at a frequency lag close to 100 $\mu$Hz equalling the large separation and 50 $\mu$Hz equalling half the value of the large separation. At 50 $\mu$Hz a clear double peak is seen  which is a signature of the small separation.} 
\end{figure}
The autocovariance spectrum of the acoustic spectrum is defined as a function of the frequency lag $L$:
\begin{equation}
A(L)=\frac{1}{N}\sum_{\nu=\nu_{\rm max}-6\Delta\nu}^{\nu=\nu_{\rm max}+6\Delta\nu}{(P_{\nu}-\bar{P})(P_{\nu+L}-\bar{P})},
\end{equation}
where $P_{\nu}$ is the value of the power density spectrum taken at the frequency $\nu$, $\bar{P}$ is the mean value of the power density spectrum in the region of interest covering $N$ frequency bins. 

In order to model the observed autocovariance spectrum we have to build a model of the power density spectrum in the region of interest and then calculate the autocovariance spectrum of this model. This model of the power density spectrum is basically identical to the model used by Fletcher et al. (2006). As free parameters we thus have the large and small frequency separations, one mean rotational frequency splitting and inclination, $\varepsilon$ which takes into account the sensitivity to the surface layers, and one mean signal-to-noise ratio of the oscillation modes. The modeling is done using the general concept of Markov chain Monte Carlo as discussed by Campante et al (2010) which allows us to provide a robust estimate of the uncertainties. We convolve the power density spectra with the observed window functions before we calculate the autocovariance spectrum. Also, in order to enhance the visibility of the features in the autocovariance spectrum we smooth the power density spectrum with a Gaussian running mean with a width of 1 $\mu$Hz (this is done both to the observed spectrum and to the model). An example of a modeled autocovariance spectrum of the acoustic spectrum of a simulation of KIC~10454133 is shown in Fig.~3. The analysis of the autocovariance spectrum of the acoustic spectrum provides us with estimates of not only $\Delta \nu$, $\delta \nu$ and $\nu_{\rm max}$, but also the mean rotational rate and inclination (see Campante et al. 2010 for a discussion of the precision with which these parameters can be obtained). For the method presented here we are mainly interested in estimates of $\Delta \nu$, $\delta \nu$ and $\nu_{\rm max}$ as these are the parameters we need for calculating the fundamental stellar parameters. Figs.~4 \& 5 show the density probability functions for $\Delta \nu$ and $\delta \nu$ for the simulation of KIC~10454133.
\begin{figure}
\centering
\includegraphics[width=5cm]{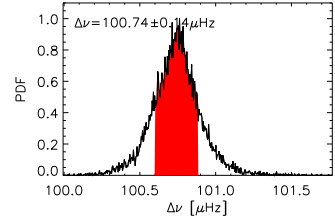}
\caption{Probability density function of the large frequency separation $\Delta \nu$ for the simulation of KIC~10454133. The red region shows the one $\sigma$ most likely solution.} 
\end{figure}
\begin{figure}
\centering
\includegraphics[width=5cm]{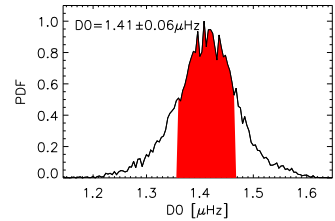}
\caption{Probability density function of the small frequency separation parameter $D_0$ for the simulation of KIC~10454133. The red region shows the one $\sigma$ most likely solution.} 
\end{figure}
\section{Stellar fundamental parameters}
We use the SEEK routine developed by P.-O.~Quirion to estimate stellar parameters based on the asteroseismic parameters $\Delta \nu$, $\delta \nu$ and $\nu_{\rm max}$ and stellar atmospheric parameters $T_{\rm eff}$, log $g$ and [Fe/H] obtained from the Kepler Input Catalog (Latham et al. 2005). SEEK also utilizes a Bayesian approach to provide robust estimates of the most likely stellar model and the uncertainties on the stellar parameters. These parameters are radius, mass, heavy element abundance, helium abundance, mixing length parameter, luminosity, surface gravity, effective temperature and age (Quirion, in prep.). Because the seismic and stellar atmospheric parameters have been obtained in a consistent manner the stellar parameters that we obtain are also consistent and possess consistent uncertainties. 
\section{Application to ground-based observations}
\begin{figure}
\centering
\includegraphics[width=6cm]{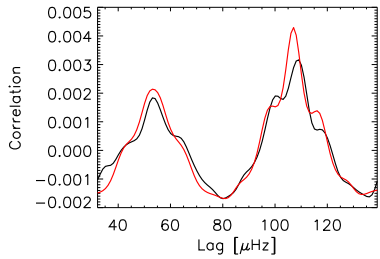}
\caption{Autocovariance spectrum of the acoustic spectrum of the ground-based observations of $\alpha$ Cen A overlaid by the model in red. The spectrum shows the signature of the small frequency separation as also see in the simulation of KIC~10454133.} 
\end{figure}
As the tools presented here, especially the one modeling the autocovariance spectrum of the acoustic spectrum, takes into account the window function they are well suited for analyzing ground-based observations which often suffer from daily gaps. We have therefore collected
time-series of high-precision radial-velocities of 10 solar-type stars ($\alpha$ Cen A \& B, $\beta$ Hydri, Procyon, $\nu$ Indi, $\tau$ Ceti, $\mu$ Arae, $\gamma$ Paw, $\delta$ Eri, HD~63077 and $\iota$ Hor) observed with HARPS and UVES from the ESO Advanced Data Products Query Form. We are in the process of analyzing these stars in order to measure in a consistent manner fundamental stellar parameters such as mass, radius, luminosity, effective temperature, surface gravity and age with consistent uncertainties. Fig.~6 shows the modeled autocovariance spectrum of the power spectrum of one of these stars $\alpha$ Cen A (Bedding et al. 2003) and Figs.~7~\&~8 show the density probability functions for $\Delta \nu$ and $\delta \nu$ for this star. Using these measurements together with stellar atmospheric parameters from Cayrel de Strobel et al. (2001) we obtain $R/{\rm R_{\odot}}=1.21\pm0.03$, $M/{\rm M_{\odot}}=1.05\pm0.09$, log$L/{\rm L_{\odot}}=0.14\pm0.02$, log $g$=4.30$\pm$0.01, $T_{\rm eff}=5760\pm100$ K and $age$=6.95$\pm$2.47 Gyr for this star which is in general agreement with the models made by Miglio \& Montalb{\'a}n (2005).
\section*{Acknowledgments}
We acknowledge the International Space Science Institute (ISSI). CK also acknowledges financial support from the Danish Natural Sciences Research Council.  

\begin{figure}
\centering
\includegraphics[width=5cm]{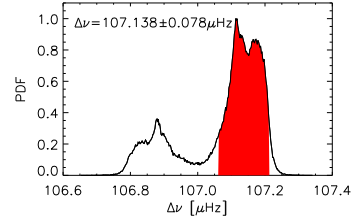}
\caption{Probability density function of the large frequency separation $\Delta \nu$ from the ground-based observations of $\alpha$ Cen A. The red region shows the one $\sigma$ most likely solution.} 
\end{figure}
\begin{figure}
\centering
\includegraphics[width=5cm]{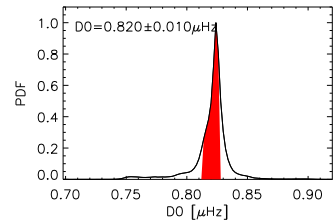}
\caption{Probability density function of the small frequency separation parameter $D_0$ from the ground-based observations of $\alpha$ Cen A. The red region shows the one $\sigma$ most likely solution.} 
\end{figure}
\end{document}